# Advancing Digital Accessibility: Integrating AR/VR and Health Tech for Inclusive Healthcare Solutions

**Vishnu Ramineni**
Albertsons Companies, Texas, USA
**Shivareddy Devarapalli**
IEEE Senior Member, Texas, USA
**Balakrishna Pothineni**
IEEE Senior Member, Texas, USA
**Prema Kumar Veerapaneni**
IEEE Senior Member, Texas, USA
**Aditya Gupta**
IEEE Senior Member, Seattle, USA
**Pankaj Gupta**
IEEE Senior Member, North Carolina, USA

**Abstract:** *Modern healthcare domain incorporates a feature of digital accessibility to ensure seamless flow of online services for the patients. However, this feature of digital accessibility poses a challenge particularly for patients with disabilities. To eradicate this issue and provide immersive and user-friendly experiences, evolving technologies like Augmented Reality (AR) and Virtual Reality (VR) are integrated in medical applications to enhance accessibility. The present research paper aims to study inclusivity and accessibility features of AR/VR in revolutionizing healthcare practices especially in domains like telemedicine, patient education, assistive tools, and rehabilitation for persons with disabilities. The current trends of advancements and case studies are also analyzed to measure the efficacy of AR/VR in healthcare. Moreover, the paper entails a detailed analysis of the challenges of its adoption particularly technical limitations, implementation costs, and regulatory aspects. Finally, the paper concludes with recommendations for integrating AR/VR to foster a more equitable and inclusive healthcare system and provide individuals with auditory, visual, and motor impairments with digital healthcare solutions.*
**Keywords:** Digital Accessibility, Americans with Disabilities Act (ADA), Web Content Accessibility Guidelines (WCAG), Augmented Reality (AR), Virtual Reality (VR), Healthcare Tech

# INTRODUCTION

Granting digital accessibility to all users irrespective of the differences such as gender, caste, and even disabilities for that matter is an immensely important feature of modern medical services. Technological advancements in the field of healthcare have transformed the delivery landscape of healthcare services, however, there remains accessibility challenges that bar the efficacy of modern digital health solutions for persons with auditory, visual, and motor disabilities (Vishnu Ramineni et al., 2025). There is persistent discrepancy in traditional digital interfaces in accommodating varied needs of users and thus, creates gap in accessing healthcare services.

However, to eradicate such disparities in healthcare access and provide immersive and user-friendly experiences, evolving technologies like Augmented Reality (AR) and Virtual Reality (VR) are integrated in medical applications to enhance accessibility. These robust technologies offer a wide range of affirmative experiences to users in their endeavor of navigating medical insights, refining rehabilitation therapies, and facilitating remote healthcare services (Ramineni et al., 2024). Through the integration of AR/VR with prevalent healthcare technologies including telemedicine platforms, wearable health monitoring devices, healthcare providers can cultivate a congenial environment of inclusivity and equality in healthcare world, especially benefitting persons with disabilities.

Nevertheless, the integration of AR/VR with prevalent healthcare technologies can result in potential challenges such as technical limitations, implementation costs, and regulatory aspects, specialized hardware requirement etc. (Bell et al, 2024). This research paper aims to study inclusivity and accessibility features of AR/VR in revolutionizing healthcare practices along with highlighting the current trends of advancements, challenges, and prospects. Also, the paper offers insights on the recommendations for integrating AR/VR to foster a more equitable and inclusive healthcare system, providing digital healthcare solutions to individuals with auditory, visual, and motor impairments.

**Applications of AR/VR in Inclusive Healthcare Solutions**

The adoption of AR/VR in healthcare has transformed the methods and ways of accessibility by providing robust solutions that boost patient engagement, therapy techniques, and medical training. Persons with disabilities also gain seamless experiences as these technologies improve healthcare delivery and patient outcome (Musamih et al., 2021). The following passages elucidate main applications of AR/VR in high accessible healthcare services. The mind map for inclusive healthcare depicted in Figure 1.

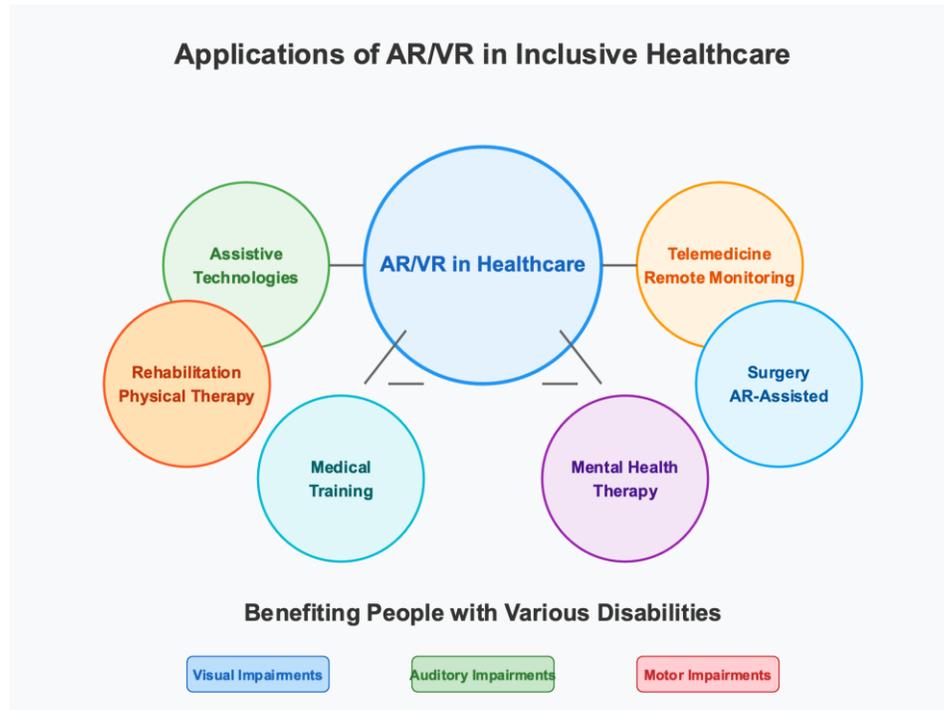

**Figure 1.** Applications of AR/VR in Inclusive Healthcare

**Assistive Technologies for Patients with Disabilities**

Individuals with auditory, visual, and motor impairments are highly benefitted by AR and VR. For instance, in case of the visually impaired users, object recognition and auditory feedback features of AR-driven applications assist them in exploring hospital environment and make interactions with digital interfaces (Ramineni et al., 2024). And in case of the users with motor disabilities, VR-driven applications help them get virtual therapy sessions which increase their mobility and motor skills (Bell et al., 2024).

**Telemedicine and Remote Patient Monitoring**

Progression in AR/VR leads to improvement in telemedicine platforms. On the one side, VR helps individuals with mobility impairments by providing them with simulated environment to interact with doctors and gain excellent telehealth experiences along with improved accessibility (Bell et al., 2024). On the other hand, AR facilitates remote consultations by overlaying patient data in real-time and help healthcare providers make informed decisions (Vishnu Ramineni et al., 2025).

**Medical Training and Education**

The medical training and education system for healthcare professionals have also witnessed significant transformation. VR simulations generate error and risk-free environment and assist healthcare providers in honing their skills before serving a real patient, along with guaranteeing accessibility in healthcare domain (Musamih et al., 2021). Again, AR applications help medical students gain complex procedural insights by offering real-time overlays of anatomical structures (Bell et al., 2024).

**Mental Health and Cognitive Therapy**

By providing controlled virtual environments for cognitive therapy, VR has proved its significance in mental health treatment scenarios. The patients suffering with Post Traumatic Stress Disorder (PTSD), anxiety, depression, phobias, and other mental disorders are given access to therapy in a protected, customizable VR setting (Ramineni et al., 2024). The rise of remote healthcare solutions during major public health disruptions has further demonstrated the need for accessible and adaptive virtual therapy platforms (Gupta et al., 2025). Besides, AR-driven applications enable a mindful environment where patients can interact with healthcare professionals and ensure their mental wellbeing (Vishnu Ramineni et al., 2025).

**Rehabilitation and Physical Therapy**

AR/VR-based applications have augmented rehabilitation programs designed to provide patients with seamless and interactive therapy sessions based on the requirements of each patient. AR wearable devices help healthcare professionals especially physiotherapists in monitoring patients' health status and offering them remedial feedback in real-time (Bell et al., 2024). Additionally, VR-based exercises create engaging and efficient therapy sessions for survivors of stroke and individuals recovering from musculoskeletal injuries by engaging them in gamified rehabilitation (Bell et al., 2024).

**Surgery and Augmented Reality-Assisted Procedures**

By incorporating AR in surgical processes, there has been improved accessibility and better precision in healthcare. The usage of AR overlays for real-time visualization of critical anatomical structures by medical surgeons has potentially minimized the risk of errors and discrepancies (Musamih et al., 2021). Again, VR-oriented preoperative simulations increase accuracy and efficacy level by helping healthcare professionals streamline complex operation processes, resulting in successful outcomes (Vishnu Ramineni et al., 2025).

**Implementing Digital Accessibility in AR/VR and Health Tech**

The approach that will guarantee digital accessibility through the integration of AR and VR technologies in healthcare for persons with disabilities must be well-organized and neatly structured. Through the integration of inclusive design principles, assistive technologies, adaptive interfaces and regulatory compliance, AR/VR technologies can potentially enhance accessibility in healthcare (Vishnu Ramineni et al., 2025). The following passages explicate fundamental strategies for employing digital accessibility in AR/VR-driven healthcare technologies. The AR/VR solutions for different types of disabilities in mentioned in Table I.

## AR/VR Solutions for Different Types of Disabilities

| Disability Type | AR Solutions | VR Solutions | Benefits |
|---|---|---|---|
| **Visual Impairments** | • Object Recognition<br>• Auditory Navigation<br>• High-Contrast Displays<br>• Text-to-Speech | • Haptic Feedback<br>• Spatial Audio<br>• Tactile Interfaces<br>• Voice Control | • Enhanced Spatial Navigation<br>• Increased Independence<br>• Better Information Access |
| **Auditory Impairments** | • Speech-to-Text<br>• Visual Notifications<br>• Captioning<br>• Visual Cues | • Sign Language Virtual Avatars<br>• Visual Interaction<br>• Haptic Alerts | • Improved Communication<br>• Enhanced Telehealth<br>• Accessible Training |
| **Motor Impairments** | • Motion Tracking<br>• Gesture Recognition<br>• Voice Commands<br>• Eye Tracking | • Virtual Therapy<br>• Gamified Exercises<br>• BCI Interfaces<br>• Adaptive Controls | • Enhanced Rehabilitation<br>• Increased Mobility<br>• Improved Independence |
| **Cognitive Impairments** | • Simplified Interfaces<br>• Step-by-Step Guides<br>• Visual Instructions<br>• Reminder Systems | • Controlled Environments<br>• Memory Training<br>• Anxiety Management | • Mental Health Support<br>• Cognitive Therapy<br>• Daily Living Skills |

**Tabel 1. AR/VR Solutions for Different Types of Disabilities**

**Inclusive Design and User-Centered Development**

Creating AR/VR applications through inclusive design ensures efficient interaction with digital health solutions for persons with disabilities. A user-oriented developmental methodology facilitates the acquisition of information from diverse populations, incorporating those with auditory, visual, and motor disabilities, to personalize AR/VR experiences as per individual needs (Bell et al., 2024). The systematic implementation of high-contrast visual displays, voice-controlled navigation protocols, and adaptable typographic scales protocols facilitate inclusivity and usability for persons with visual disabilities (Ramineni et al., 2024).

**Assistive Technologies and Adaptive Interfaces**

Assistive technological modalities that integrate screen readers, vocal commands, and gestural interfaces boost accessibility in AR/VR healthcare applications. Through illustration, AR-driven smart spectacles provide object identification and auditory directives for persons with visual impairments, while VR-enabled rehabilitation platforms make use of motion-capture process to assist persons with mobility deficits (Bell et al., 2024). Moreover, adaptive interfaces augment accessibility and usability through modifications in display parameters or interaction modalities (Musamih et al., 2021).

**Compliance with Digital Accessibility Standards**

One of the primary prerequisites for ensuring inclusivity within the domain of AR/VR health applications is to adhere to distinct digital accessibility guidelines. Well-defined standards like Web Content Accessibility Guidelines (WCAG), Section 508 of the Rehabilitation Act along with the Americans with Disabilities Act (ADA) form systematic frameworks for the digital accessibility (Ramineni et al., 2024). Healthcare institutions that incorporate AR/VR technologies must evaluate adherence level of accessibility and usability to certify conformity with these given standards (Bell et al., 2024).

**Enhancing AR/VR for Telemedicine and Remote Healthcare**

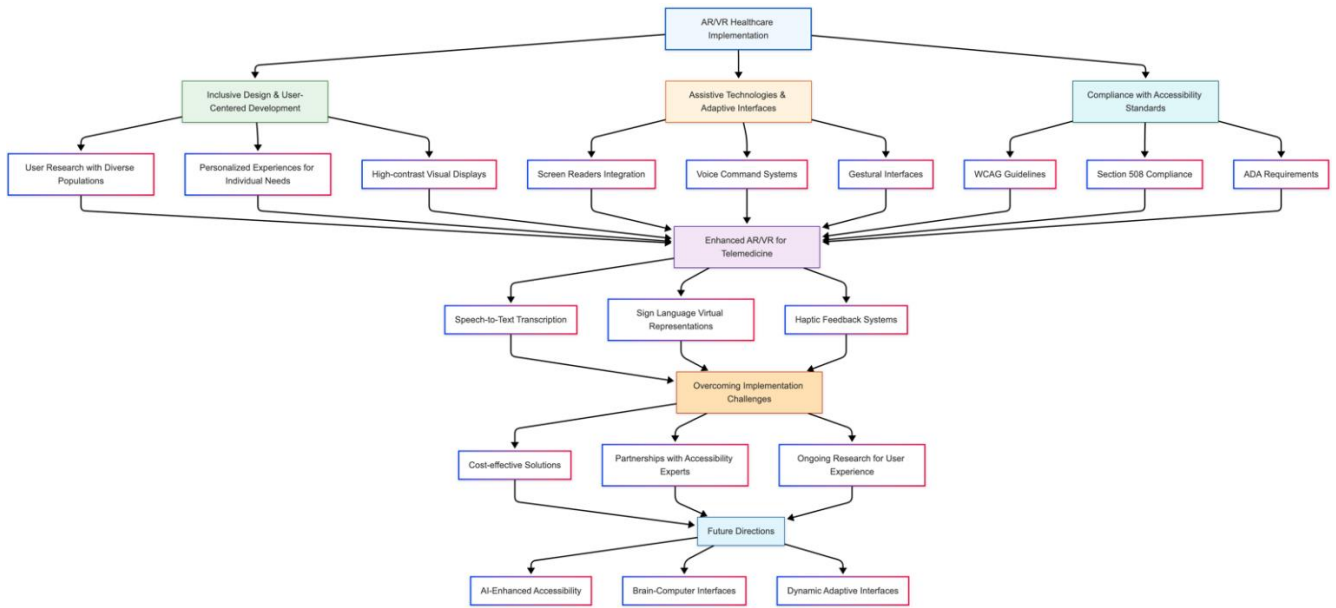

**Figure 2.**                  Flow chart for enhancing accessibility in AR/VR for telemedicine

Incorporation of AR/VR technologies serves as a pivotal way to boost accessibility of telemedicine through dynamic and synchronous healthcare interactions. Digital medical consultation spaces facilitating speech-to-text transcription, sign language-assisted virtual representations, and haptic feedback procedures, enable efficacious communication between persons with auditory or speech disabilities and healthcare providers (Bell et al., 2024). Moreover, AR-driven technologies in healthcare can help medical professionals gain patients' medical insights in real-time, thereby ensuring remote patients' diagnosis with ease and clarity (Ramineni et al., 2024). The methodology for enhancing accessibility is described in Figure 2.

**Overcoming Challenges in AR/VR Accessibility Implementation**

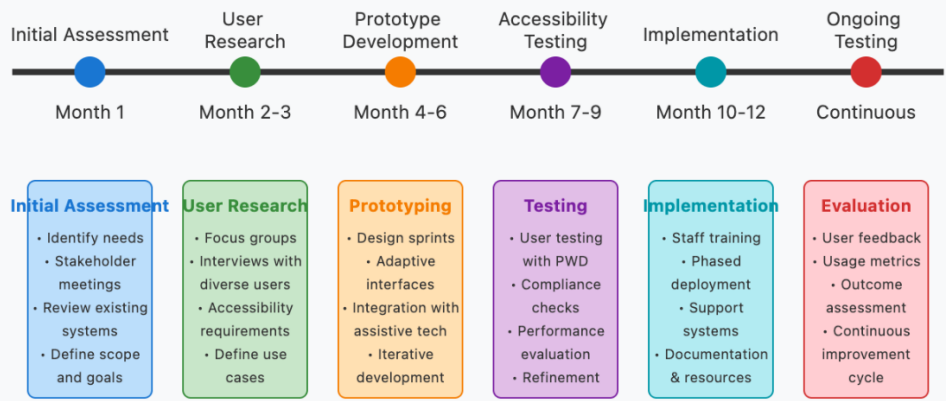

Figure 3.Timeline for AR/VR accessibility implementation in Healthcare

Notwithstanding the fundamental pros of AR/VR within healthcare domain, many barriers restrict their extensive implementation. Some of the basic impediments to global accessibility include high implementation expenditures, the requirement for specialized technological infrastructure, and inherent technical restrains, such as kinetosis and cognitive saturation (Musamih et al., 2021). Efficiently eradicating these barriers means systematized investment in economically viable solutions, collaborative partnerships with specialists in accessibility, and sustained scholarly inquiry focused on the progressive increase of user experience across diverse populations with disabilities (Bell et al., 2024). The timeline for accessibility implementation is mentioned in Figure 3.

**Future Directions for AR/VR Accessibility in Healthcare**

Advancement in the realm of artificial intelligence (AI) and machine learning (ML) is proportionately related to amplifying the accessibility of AR/VR within healthcare contexts. AI-enabled voice-operated assistants, instantaneous language translation services, and AI-driven dynamically adjusting interfaces can potentially improve accessibility for persons with diverse ipairments (Ramineni et al., 2024). Additionally, the systematic integration of AR/VR technologies

81

with brain-computer interfaces (BCIs) may unveil new interaction models for persons with motor impairments, affording them the capability to interact with healthcare applications via neural signals (Ramineni et al., 2024).

**Enhancing Digital Accessibility in AR/VR and Health Tech**

Digital accessibility is highly essential in healthcare domain as it ensures that everyone irrespective of their differences, abilities, and interests, can easily utilize these AR/VR-assisted applications. Assistive technologies are integrated, regulations are sincerely adhered to, and inclusivity is ensured through digital accessibility features in AR/VR-based technologies (Ramineni et al., 2025). The following sections elucidate the strategies necessary to increase accessibility in AR/VR healthcare applications. Distribution of key accessibility features in AR/VR Healthcare is depicted in Figure 4.

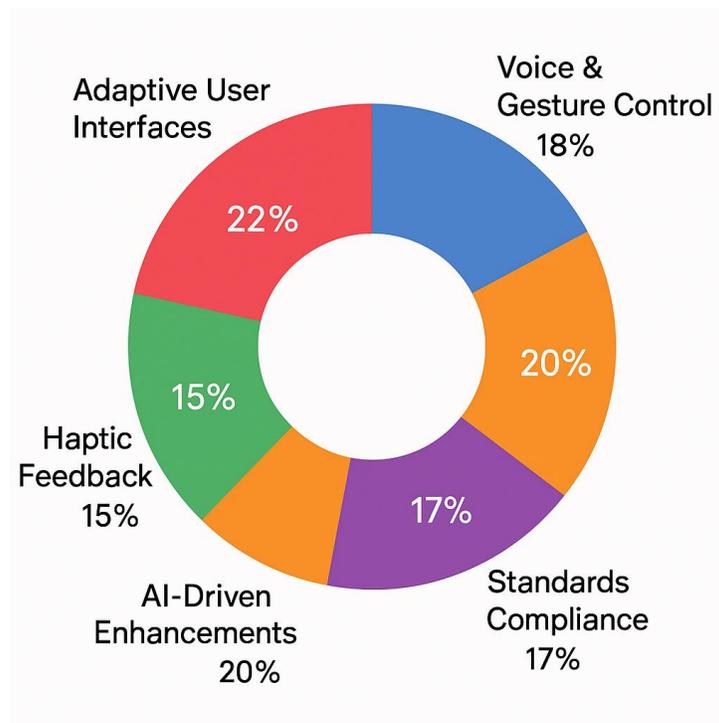

**Figure 4.** Distribution of accessibility features in AR/VR for Healthcare

**Adaptive User Interfaces and Personalization**

An important strategy to ensure digital accessibility is related to deployment of adaptive user interfaces that are strategically tailored to the personal requirements of users. Thus, AR/VR healthcare applications must involve personalized features such as text-speech options, adaptable fonts, high-contrasts themes, and gesture regulating options (Musamih et al., 2021), which assist persons with auditory, visual, motor disabilities, thereby ascertaining their efficient interaction with these robust technological devices (Ramineni et al., 2024). Personalization of healthcare interfaces based on context and user state has shown to improve accessibility outcomes (Gusain et al., 2017).

**Voice and Gesture-Controlled Interaction**

To further increase digital accessibility in AR/VR healthcare, tech developers must put emphasis on incorporating multimodal interaction techniques that prioritizes gesture, voice, and eye-tracking inputs, thereby creating flexible environment for individuals with disabilities to interact with these applications. Moreover, integration of haptic feedback can offer vital sensory insights for users with visual impairments, while guaranteeing compatibility with prevalent assistive technologies like screen readers and switch devices is pivotal for a flawless experience of inclusivity. Therefore, a user-oriented design approach that incorporates frequent feedback from persons with disabilities will eventually help in the formation of effective, equitable, and accessible AR/VR healthcare solutions.

**Haptic Feedback and Multisensory Accessibility**

The integration of haptic feedback constitutes a vital aspect in increasing the accessibility of AR/VR healthcare technologies for persons with sensory disabilities. What enhances user navigation within virtual settings are the tactile response modalities which consist of vibrations and force feedback and thus, increases their capacity to interact with digital health assets (Ramineni et al., 2024). Moreover, in case of persons with multiple disabilities, the accessibility of AR/VR application is ensured through multisensory integration and combination of auditory, visual, and haptic cues (Ramineni et al., 2024).

**Real-Time AI-Driven Accessibility Enhancements**

With the progression in Artificial Intelligence (AI), real-time accessibility is enhanced in the AR/VR healthcare solutions. AI-assisted text-to-speech and speech-to-text systems improves the communication process for persons with auditory and speech disabilities (Musamih et al., 2021). Again, for users with visual disabilities, AI-driven image recognition feature assists them by offering auditory descriptions of objects and surrounding in AR-augmented applications (Bell et al., 2024). These AI-enabled solutions potentially enhance usability and accessibility in digital healthcare domain.

**Compliance with Global Accessibility Standards**

Adherence to global accessibility standards is pivotal for ensuring inclusivity and equality in AR/VR-based healthcare solutions. Well-established frameworks like Web Content Accessibility Guidelines (WCAG), Section 508 of the Rehabilitation Act along with the Americans with Disabilities Act (ADA) must be complied with by the AR/VR-driven health application developing organizations (Bell et al., 2024). General user testing encompassing data of diverse population and compliance audits aid in the identification of accessibility gaps and help ascertain inclusivity and equality in digital healthcare solutions (Vishnu Ramineni et al., 2025).

**Future Trends in AR/VR Accessibility**

The expanding growth in machine learning, neurotechnology, and adaptive computing shapes the future of digital accessibility. Brain-computer interfaces (BCIs) proves as one of the accessibility solutions that potentially support individuals with motor disabilities to use neural signals while interacting with AR/VR applications (Ramineni et al., 2024). Besides, wearable assistive devices like smart glasses and biometric sensors which incorporates AR/VR technologies further boost healthcare accessibility for persons with impairments (Bell et al., 2024).

**Challenges in Digital Accessibility for Integrating AR/VR and Health Tech**

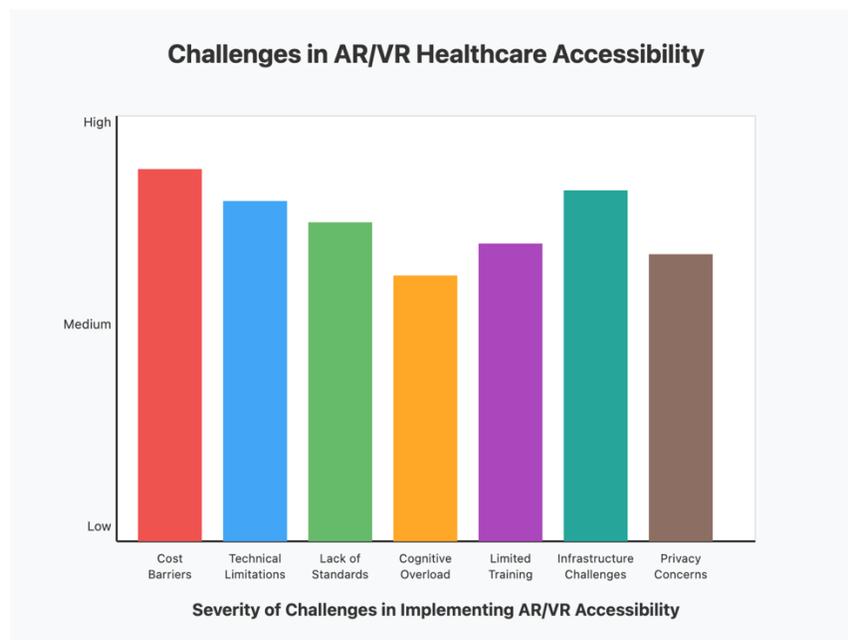

Figure 5.Challenges in AR/VR Healthcare Accessibility

AR/VR technologies provide exclusive avenues to improve accessibility in healthcare domain, yet potential impediments obstruct their widespread and efficient use, especially for persons with disabilities. Some of the major barriers are related to technical limitations, substantial costs, absence of uniform standards, and issues with users' adaptability (Vishnu Ramineni et al., 2025). To overcome these barriers is primarily significant to realize the promise of inclusivity through integration of AR/VR technologies and guarantee their aids reach every user as mentioned in Figure 5.

**High Cost of Implementation**

The high implementation cost for outlaying accessibility features in AR/VR-based healthcare products seem to be one of the key barriers. Since the price tags for premium AR-driven smart glasses, VR headsets, and robust haptic devices often deter healthcare amenities as well as the patients (Ramineni et al., 2024). Additionally, the customization requirements to comply with the accessibility guidelines substantially increases development expenses, making such technological integration particularly inconvenient for smaller healthcare organizations (Bell et al., 2024).

**Technical Limitations and Hardware Constraints**

Notwithstanding the technological progressions, AR/VR hardware carry several limitations that poignantly hamper accessibility. In case of users with vestibular disorders, VR headsets lead to motion sickness (Bell et al., 2024). Moreover, several devices are devoid of built-in accessibility features including eye-tracking, voice control, and adaptive interfaces, thereby making it inconvenient for persons with disabilities to manage and interact with these technologies efficiently (Musamih et al., 2021)

**5.3. Lack of Standardized Accessibility Guidelines**

It's to be noted that web accessibility is guided by frameworks like Web Content Accessibility Guidelines (WCAG) but VR/AR accessibility are devoid of any uniformly standardized guidelines (Vishnu Ramineni et al., 2025). This lack of universally acknowledged guideline results in inconvenient for developers to frame and implement standard accessibility features across varied VR/AR platforms (Bell et al., 2024). Such a limitation heightens the risk of non-adherence to disabilities laws, and thus limits the integration of accessible healthcare technologies.

**Cognitive and Sensory Overload**

There is high possibility in AR/VR environment becoming excessively overwhelming with auditory, visual, and interactive stimuli, making it difficult for users with disabilities to interact with these technologies. For individuals with neurodiverse conditions including autism spectrum disorder, these overwhelming flow of insights/follow-ups, unnaturally bright light, complex interactive process create a distressful environment and deter their usability (Bell et al. 2024).

**Limited User Awareness and Training**

Healthcare professionals and patients lack adequate training, knowledge, and expertise of how AR/VR technologies are adapted to enhance accessibility. This lack of training and knowledge ultimately result in ineffective utilization of these sophisticated technologies (Musamih et al. 2021). Moreover, what bars persons with disabilities to independently set up and utilize VR/AR healthcare applications is the lack of accessible user manuals and supervisory materials (Vishnu Ramineni et al., 2025).

**Connectivity and Infrastructure Challenges**

High-speed internet connectivity and proper infrastructure facility boosts the performativity and effectiveness of AR/VR applications. Additionally, the computing devices must be powerful as well. Therefore, in areas with inadequate bandwidth, improper technological infrastructure, inadequate network latency, and outdated hardware, the accessibility is highly constrained (Bell et al., 2024). The residents of rural or underserved areas encounter difficulty in getting access to AR/VR-driven healthcare services owing to these limitations (Bell et al., 2024).

**Ethical and Privacy Concerns**

Sensitive medical data of users are collected and processed by AR/VR-assisted healthcare platforms or applications. To guarantee that these technologies duly adhere to the privacy regulations like Health Insurance Portability and Accountability Act (HIPAA) and the General Data Protection Regulation (GDPR) seems to be immensely challenging (Ramineni et al., 2024). It's to be noted that users' confidentiality and data privacy can be compromised easily through unauthorized access to AR/VR-based technologies. This in turn leads to ethical consideration about user data security and tracking (Vishnu Ramineni et al., 2025)

**Future Directions in Overcoming Accessibility Challenges**

To effectively overcome these accessibility limitations, AR/VR healthcare applications, for future use, must ensure cost-effective approach, better hardware adaptability, and compliance with standardized accessibility guidelines. Policymakers, healthcare practitioners, and technology developers must collaborate in guaranteeing the VR/AR-assisted applications are carved out while keeping inclusivity and equal accessibility in mind (Bell et al., 2024). Moreover, through integration of Artificial Intelligence (AI) and Machine Learning (ML), there can be improved accessibility with seamless real-time speech recognition feature, proficiently adaptive user interfaces, error-free gesture-based interactions (Bell et al., 2024). Similar predictive frameworks in adaptive systems have demonstrated success in structured environments (Gupta et al., 2016).

**CONCLUSION**

With the adoption of Augmented Reality (AR) and Virtual Reality (VR) in healthcare, digital accessibility especially for persons with disabilities has been potentially transformed. There has been improvement of patient rehabilitation, medical training, telemedicine, and assistive solutions for persons with visual, auditory, and mobility impairments, as the patients are exposed to immersive experiences (Vishnu Ramineni et al., 2025). Nevertheless, in addition to these benefits, some potential barriers including hardware limitations, high implementation costs, no standard guideline of accessibility, and privacy breach matters pose serious challenges, thereby obstructing its adoption (Ramineni et al., 2024).

Implementation of cost-effective AR/VR solutions along with setting up accessibility standards and ensuring adaptability of device can meaningfully help in the eradication of these impediments. Another prospect can be ascertaining due collaboration between tech developers, policymakers, and healthcare practitioners to allow for an accessible and equitable healthcare environment where individuals with disabilities do not encounter biases or ignorance, and thus, ensuring alignment of these technologies with standard accessibility regulations (Bell et al., 2024). Moreover, through integration of Artificial Intelligence (AI) and Machine Learning (ML), there can be improved accessibility with seamless real-time speech recognition feature, proficiently adaptive user interfaces, error-free gesture-based interactions (Bell et al., 2024).

The unceasing growth of Augmented Reality and Virtual Reality technologies enhances its predominant value in healthcare industry primarily due its accessibility feature. These technologies outgrow other technologies as a powerful tool for addressing key challenges, cultivating equitable healthcare ecosystem, and enabling seamless experience through inclusivity. For persons with auditory, visual, or motos impairments, the integration of AR/VR innovations in healthcare is a boon as it advances digital accessibility and create a more inclusive and patient-centric healthcare approach.